\documentclass[prl,aps,twocolumn,showpacs,superscriptaddress,nofootinbib]{revtex4-1}
\pdfoutput=1
\nonstopmode
\usepackage{amssymb}
\usepackage[english]{babel}
\usepackage[utf8]{inputenc}
\usepackage{amsfonts}
\usepackage{amsmath}
\usepackage{graphicx}
\usepackage{bm}
\usepackage{color}
\usepackage{pdfpages}

\newcommand{\nohat}{}

\begin{document}


\title{Optimal Control of Open Quantum Systems: Cooperative Effects of Driving and Dissipation}


\newcommand{\aft}{Institut f\"ur Theoretische Physik, Universit\"at
Ulm, Albert-Einstein-Allee 11, 89069 Ulm, Germany.}
\newcommand{\afq}{Institut f\"ur Quanteninformationsverarbeitung,
Universit\"at Ulm, Albert-Einstein-Allee 11, 89069 Ulm, Germany.}

\author{R. Schmidt}\affiliation{\aft}
\author{A. Negretti}\affiliation{\afq}
\author{J. Ankerhold}\affiliation{\aft}
\author{T. Calarco}\affiliation{\afq}
\author{J. T. Stockburger}\affiliation{\aft}

\date{August 22, 2011}


\begin{abstract}
We investigate the optimal control of open quantum systems, in
particular, the mutual influence of driving and dissipation.  A
stochastic approach to open-system control is developed, using a
generalized version of Krotov's iterative algorithm, with no need for
Markovian or rotating-wave approximations. The application to a
harmonic degree of freedom reveals cooperative effects of driving and
dissipation that a standard Markovian treatment cannot
capture. Remarkably, control can modify the open-system dynamics to the point
where the entropy change turns negative, thus achieving cooling of
translational motion without any reliance on internal degrees of
freedom.
\end{abstract}



\maketitle


\emph{Introduction}.  The control of quantum dynamics for the accurate
preparation of a prescribed quantum state by a tailored time-dependent
field is a task of key importance in quantum physics and related
disciplines. With increasing complexity of devices for quantum
information processing the destructive role of environmental
fluctuations has become a severe limitation to further progress.  For
example, neutral atoms or ions in electromagnetic traps are exposed to
fluctuations of (comparatively hot) chip surfaces~\cite{treutlein04},
while in superconducting circuits diffusing charges and
electromagnetic fluctuations affect fidelities quite substantially
\cite{ithie05}.

In this context, optimal control theory (OCT) has emerged as a key
ingredient in strategies to tame the effect of decoherence and other
imperfections either directly by mitigating their effect or indirectly
by speeding up operations. In limiting cases or under idealized
conditions OCT has already been used to increase the fidelity of
simple quantum
gates~\cite{montangero07,tesch02,sporl07,schirmer09,palao02} or to
construct more complex protocols~\cite{schulte05}. Generally, however,
OCT has treated environmental interactions mostly by heuristic or
approximate methods so far. A simple strategy has been followed in
Ref.~\cite{treutlein06}, where the impact of a heat bath was taken
into account by assuming an initial thermal state while neglecting its
effect completely during its dynamics.  Some progress has been
achieved within fully dynamical approaches based on standard Markovian
master equations.  Unfortunately, these schemes have severe drawbacks:
First, they become inconsistent for strong control fields unless
additional field-dependent memory terms in the dissipator are
introduced~\cite{kofma04,xu04}. Second, the rotating wave
approximation (RWA), which is usually performed on the system-environment
interaction in order to obtain a master equation with complete
positivity, is known to be unreliable in driven
systems~\cite{kofma04}.  Third, at sufficiently low temperatures or
for reservoirs with structured spectral mode densities the true
dynamics of open quantum systems is non-Markovian. These errors in the
dynamics may be further amplified by OCT search algorithms. As a
result, OCT computations using standard master equations yield
mismatched fields, which perform poorly when applied to a realistic
setting.

In this Letter we present a treatment of optimal control of open
quantum systems which is not susceptible to these problems. Stochastic
Liouville-von Neumann (SLN) equations~\cite{stock02,Koch2008}
provide an approach to quantum dissipation in a driven system which is
both conceptually transparent and formally exact. Field-induced
modifications of environmental effects are thus included a
priori. From the SLN equations one arrives naturally at the definition
of a state-costate pair \cite{foot:costate} of dynamical variables
with the simple first-order equations of motion required by
gradient-based and related OCT methods.

We outline the salient features of our OCT technique (for details see
Supplemental Material \cite{supplprl11}) and apply it to a simple
model.  Notably, it turns out that control fields extracted from a
RWA-based master equation differ substantially from exact ones
obtained within the SLN scheme. This leads even to the
counterintuitive phenomenon of control-induced cooling, which is
completely missing in the RWA approach.

\emph{Open-system dynamics and control algorithm}.---We start with the
exact stochastic equation of motion \cite{stock02}
\begin{eqnarray}
\dot{\nohat{\varrho}}_{\xi,\nu}(t)
 &=&
-\frac{i}{\hbar}[\nohat{H}_{\mathrm{s}},\nohat{\varrho}_{\xi,\nu}(t)]
-\frac{i}{\hbar}[\nohat{H}_{\mathrm{c}},\nohat{\varrho}_{\xi,\nu}(t)]
\nonumber\\
&&  + \frac{i}{\hbar}\xi(t)[\nohat q,\nohat{\varrho}_{\xi,\nu}(t)]
 +\frac{i}{2}\nu(t)\{\nohat q,\nohat{\varrho}_{\xi,\nu}\}
\label{eq:SLN}
\end{eqnarray}
for noisy samples $\nohat\varrho_{\xi,\nu}(t)$ of the density matrix
of an open quantum system. The system is characterized by a
Hamiltonian $\nohat{H}_{\mathrm{s}}$ governing its autonomous motion,
augmented by a term $\nohat{H}_{\mathrm{c}}(t)$ describing the
influence of time-dependent control fields $u_j(t)$. It interacts
bilinearly with a dissipative environment whose quantum statistical
force-force correlation function is mapped to the noise statistics as
follows~\cite{stock02}: (i) the autocorrelation of $\xi$ matches the
quantum noise of the reservoir, (ii) the cross-correlations of $\xi$
and $\nu$ match the dynamical response of the environment and (iii)
the correlations $\langle\nu(t)\nu(t')\rangle$ vanish
identically. According to the last condition $\nu(t)$ is a complex
variable with random phase.  The non-vanishing correlations are
identical to the real and imaginary parts of the kernel defining the
Feynman-Vernon influence functional \cite{weiss08b,tanim06}, from
which Eq. (\ref{eq:SLN}) can be derived without approximations.

The fluctuations and the dynamical response of the thermal environment
are fully characterized by its temperature and spectral density
$J(\omega)$, which in the present case will be taken as Ohmic
(proportional to $\omega$ up to a UV cutoff $\omega_c$). Both
$1/\omega_c$ and, more important, the thermal time $\hbar\beta$ are
intrinsic time scales of the environmental fluctuations.
Eq. (\ref{eq:SLN}) now simplifies to
\begin{eqnarray}
\dot{\nohat{\varrho}}_{\xi}(t) &=& \nohat{\mathfrak{L}} \nohat{\varrho}_{\xi}(t) :=
-\frac{i}{\hbar}[\nohat{H}_{\mathrm{s}},\nohat{\varrho}_{\xi}(t)]
-\frac{i}{\hbar}[\nohat{H}_{\mathrm{c}},\nohat{\varrho}_{\xi}(t)]
\nonumber\\
&&-\frac{i}{2\hbar}\gamma_{0}[\nohat q
 ,\{\nohat{p},\nohat{\varrho}_{\xi}(t)\}]
 + \frac{i}{\hbar}\xi(t)[\nohat q,\nohat{\varrho}_{\xi}(t)]\, ,
\label{eq:SLED}
\end{eqnarray}
where $\gamma_0$ is the damping rate of a Brownian particle of mass $m$~\cite{stock99}. The physical density matrix is a
stochastic average of the form $\nohat\varrho(t) =
\mathbb{E}[\nohat\varrho_{\xi}(t)]$.

At the price of introducing an explicit noise variable $\xi(t)$,
Eq. (\ref{eq:SLED}) represents the \emph{exact non-Markovian} dynamics
in terms of a stochastic ensemble with \emph{time-local} equations of
motion. All memory effects inherent in the reservoir dynamics are
contained in the time-dependent correlations. There is no
decomposition of the environment into additional degrees of freedom
and a secondary, Markovian environment~\cite{reben09}.  In an extreme
high-temperature limit, Eq.\ (\ref{eq:SLED}) becomes Markovian and
reduces to the master equation of Caldeira and
Leggett~\cite{calde83b}.

Optimal control means searching for control signals which drive
desirable characteristics of the dynamics to extremal values.
Here we consider optimization objectives defined by minimization of an
expectation value $\langle\nohat{M}\rangle$ at a specified end time $T$.
This leads to a search for extrema of the objective
functional
\begin{equation}
\mathsf{B}[{{\bf u}}(t)] =
 \mathbb{E}[\mathop{\rm tr}\{\nohat{M} \nohat\varrho_{\xi}(T)\}]
= \mathop{\rm tr}\{\nohat{M}\nohat\varrho(T)\},
\end{equation}
where the first equality relies on the map ${\bf u} \mapsto
\nohat\varrho_{\xi}(T)$ implicit in the initial-value problem of
Eq.\ (\ref{eq:SLED}).
Alternatively, Eq.\ (\ref{eq:SLED}) can be interpreted as a constraint
on simultaneous variations of ${\bf u}(t)$ and $\nohat\varrho_{\xi}(t)$.
This constraint needs to be taken into account through a
time-dependent Lagrange multiplier $\nohat\Lambda_{\xi}(t)$, which also
depends on the particular noise realization $\xi(t)$. Variational
calculus leads to the equation of motion
\begin{eqnarray}
\dot{\nohat{\Lambda}}_{\xi}(t) &=& -
\nohat{\mathfrak{L}}^{\dagger}\nohat{\Lambda}_{\xi}(t) =
-\frac{i}{\hbar}[\nohat{H}_{\mathrm{s}},\nohat{\Lambda}_{\xi}(t)]
-\frac{i}{\hbar}[\nohat{H}_{\mathrm{c}},\nohat{\Lambda}_{\xi}(t)]
\nonumber\\
&&-\frac{i}{2\hbar}\gamma_{0}\{\nohat{p}, [\nohat q
 ,\nohat{\Lambda}_{\xi}(t)]\}
 + \frac{i}{\hbar}\xi(t)[\nohat q,\nohat{\Lambda}_{\xi}(t)]
\label{eq:SLEDadj}
\end{eqnarray}
for $\nohat\Lambda_{\xi}(t)$, called the costate of the optimal control
problem in this context. Equation (\ref{eq:SLEDadj}) is not an initial
value problem; it needs to be solved with the terminal boundary
condition $\nohat{\Lambda}_{\xi}(T) + \nohat{M}=0$ arising from
variation of the final state.

Now the gradient of the objective functional under the above constraint is given by
\begin{equation}\label{eq:update}
\left. \frac{\delta \mathsf{B}}{\delta u_j(t)}\right|_\textrm{constr.} =
 \mathbb{E}\left[
{\rm tr}\left\{ \nohat{\Lambda}_{\xi}(t)\frac{\partial
  \nohat{\mathfrak{L}}}{\partial u_{j}}\nohat{\varrho}_{\xi}(t)
\right\}
\right],
\end{equation}
where $\nohat\varrho_{\xi}(t)$ and $\nohat{\Lambda}_{\xi}(t)$ obey the
stochastic equations of motion (\ref{eq:SLED}) and (\ref{eq:SLEDadj}).

The preceding considerations show that the SLN approach treats quantum
memory effects in a mathematical language which integrates seamlessly
into the state-costate framework of standard OCT techniques.  For
numerical computations, we have adapted the monotonically convergent
algorithm of Krotov \cite{Krotov1996,Sklarz2002} to the present case
of non-Markovian stochastic propagation. This algorithm improves
performance by substituting gradient search steps with a nonlocal
generalization of Eq.\ (\ref{eq:update}). Key performance
characteristics are improved by this change (see~\cite{supplprl11}).

\emph{Application}.---As a common model we consider a harmonic
oscillator, i.e.,
$\nohat{H}_{\mathrm{s}}=\frac{\nohat{p}^{2}}{2\,m}+\frac{m
\omega_0^{2}}{2}\nohat{q}^{2}$, which is subject to an additional
potential $\nohat{H}_{\mathrm{c}}(t)=-F(t)\nohat{q}+\frac{m}{2}
\Delta(t)\nohat{q}^{2}$ depending on a force control fields $F(t)\equiv
u_1(t)$ and a tuning control field $\Delta(t)\equiv u_2(t)$, which
modifies the instantaneous resonance frequency, $\omega^2 = \omega_0^2
+ \Delta$.  This choice is not only a model for typical realizations
as, e.g.,\ trapped atoms or ions or low energy dynamics of Josephson
junctions, it also offers itself as a simple test case where an
intuitive interpretation of numerical results may be feasible.  It is
nontrivial since it includes the nonlinear response of the system to
parametric driving, and it is fairly generic as it applies to
potentials with harmonic minima.

Under the equation of motion (\ref{eq:SLED}), the individual samples
$\nohat{\varrho}_{\xi}$ remain Gaussian for Gaussian initial states.
This allows us to rephrase the equation of motion (\ref{eq:SLED}) as
a system of ordinary stochastic differential equations for the first
and second cumulants (means and variances) associated with
$\nohat\varrho_{\xi}$, i.e., $\langle \nohat{q} \rangle_{c}$, $\langle
\nohat{p} \rangle_{c}$, $\langle \nohat{q}^{2} \rangle_{c}$, $\langle
\nohat{p}^{2} \rangle_{c}$ and
$\langle\frac{1}{2}(\nohat{q}\nohat{p}+\nohat{p}\nohat{q}) \rangle_{c}$.  A
similar consideration holds for the costate dynamics
(\ref{eq:SLEDadj}) if a maximal overlap with a Gaussian target state
is chosen as optimization objective, i.e., $\nohat{M} = 1 - \nohat{A}$,
where $\nohat{A} = |\alpha\rangle\langle\alpha|$ projects onto a
coherent state. We thus obtain closed equations of motion for the
first two cumulants in the propagation of both the state
$\nohat\varrho_{\xi}(t)$ and the costate $\nohat\Lambda_{\xi}(t)$.  While
the effect of the linear control $F(t)$ alone is given
by linear response theory, the dynamical squeezing through a
time-dependent $\Delta(t)$ leads to nontrivial dynamics, as does the
combined action of both controls.  We have explored these effects
numerically, computing the expectation values in Eq. (\ref{eq:update})
explicitly through a large number of samples (typically $10^4$). This
has the advantage of being securely based on first principles, without
resorting to approximations of the dynamics.

In the following, we use natural units ($\hbar=k_{\rm B}=1$, units
$\omega_0$ for energies, angular frequencies, or rates,
$1/\sqrt{m\omega_0}$ for lengths, and $\sqrt{m\omega_0}$ for
momenta). We choose a minimal-uncertainty wavepacket centered around
$q=1$ and $p=0$ as both initial and target state. Values of the
temperature and the damping constant are chosen in the range typical
of superconducting solid-state devices~\cite{ithie05}. The propagation
time $T=20$ is roughly comparable to the relaxation time in the
examples to be discussed.

We compare the results of iteratively determined control fields for
three types of dynamics inserted for state and costate in Eq.\
(\ref{eq:update}): (a) SLN dynamics; (b) the standard Markovian master
equation of the harmonic oscillator~\cite{breue02}, with the usual
raising and lowering operators associated with $\nohat{H}_{\mathrm{s}}$
as Lindblad operators; (c) quantum dynamics without dissipation.

\begin{figure}
\begin{center}
\includegraphics[width=0.95\linewidth]{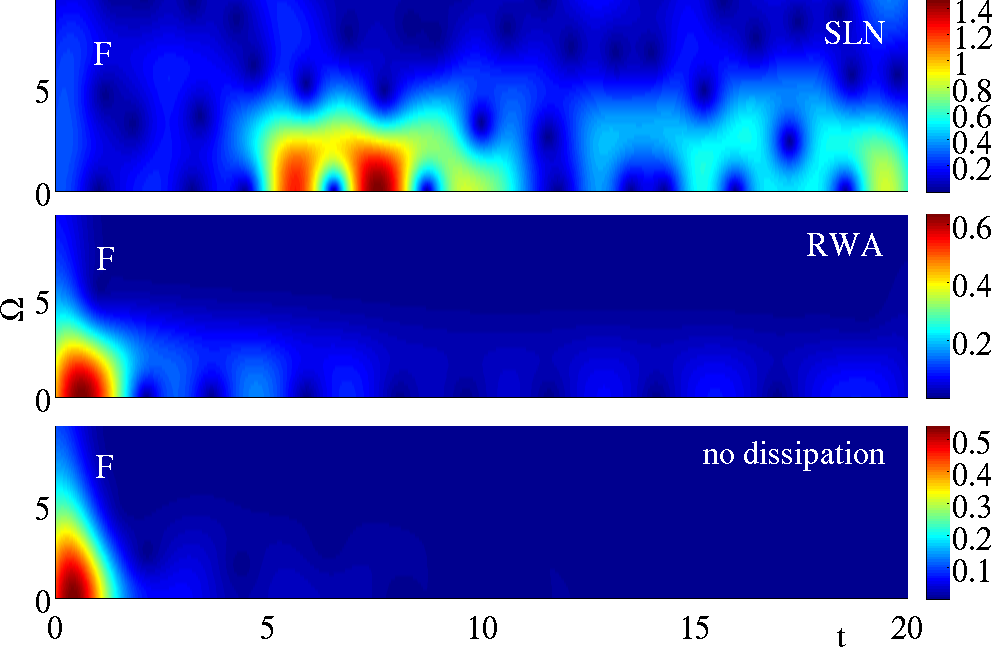}
\end{center}
\caption{(color online) Windowed Fourier transform of the optimal control
force $F(t)$ obtained using different dynamical equations: (a) SLN
equation (\ref{eq:SLED}), SLN, (b) a simple generalization of the
standard master equation to driven systems, RWA, and (c) unitary
propagation. Parameters are $\gamma_0=0.05$, $\omega_c=50$, $\beta=1$.}
\label{fig:u1}
\end{figure}

\begin{figure}
\begin{center}
\includegraphics[width=0.95\linewidth]{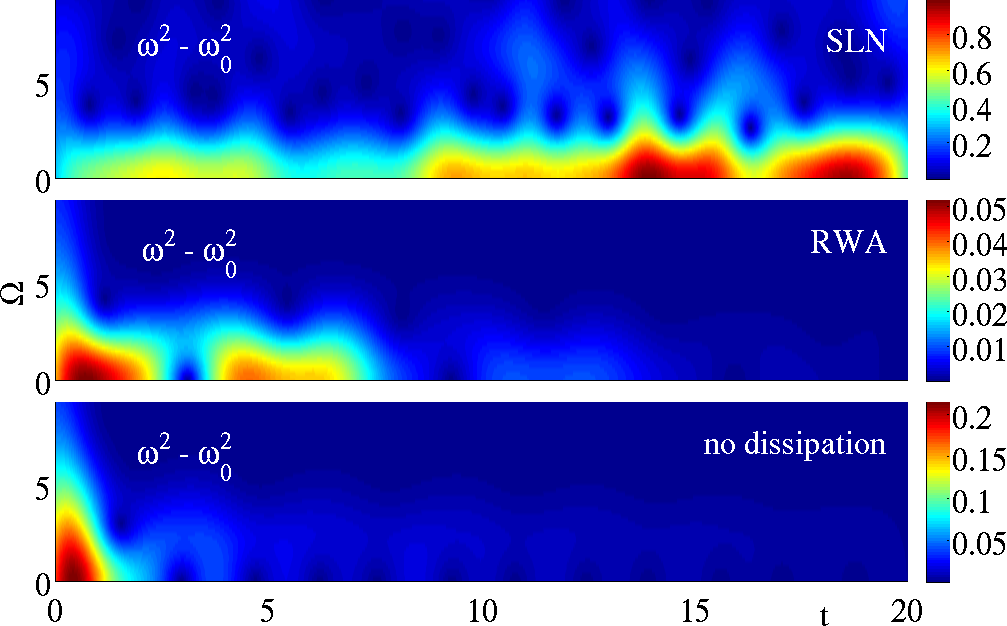}
\end{center}
\caption{(color online) Windowed Fourier transform of the optimal
tuning field $\Delta = \omega^2 - \omega_0^2$ obtained using dynamical
equations as in Fig.\ \ref{fig:u1}. Different color scales apply to
the three scenarios.}
\label{fig:u2}
\end{figure}

Figures \ref{fig:u1} and \ref{fig:u2} show time-frequency signatures
of the controls $F(t)$ and $\Delta(t)$ obtained through the windowed
Fourier transform (also short-time Fourier transform, STFT) using a
Gaussian window.  Both controls show marked differences between the
SLN and RWA cases. The tendency for more pronounced and more complex
high-frequency features in the SLN case indicates the importance of
exercising control also on time scales of the environmental
fluctuations (of order $\beta$), similar to a known strong-field
approach to the suppression of decoherence known as `bang-bang
control'~\cite{viola98}. A second tendency seen in the SLN results is
the application of fields spread out over the entire time interval, as
compared to the emphasis on a stronger initial perturbation in the
cases of RWA dissipation or no dissipation.

Values of the objective functional achieved with the SLN fields for
different temperatures and damping constants are compared in Table
\ref{tab:obj}. Free dynamics (no control) would result in values
roughly equal to $1/2$ for all parameters listed. A test of the
control fields obtained in RWA, inserted in the exact equation of
motion, typically yields values of the objective functional which are
up to 100\% larger than for controls computed using SLN dynamics. The
algorithmic property of monotonic convergence is confirmed by our
numerical results.

\begin{table}[b]
\begin{tabular}{c||p{2cm}|p{2cm}|p{2cm}}
$\beta \backslash \gamma_{0}$ & 0.005 & 0.01 & 0.05 \\ \hline\hline
$0.5$ &  0.1036 & 0.1582  & 0.3351  \\ \hline
$1.0$ & 0.0477 & 0.0688  & 0.1432  \\ \hline
$5.0$ & 0.0059 & 0.0109  & 0.0245 \\ \hline
$50.0$ &  0.0037  & 0.0072 & 0.0133 \\ 
\end{tabular}
\caption{Results for the minimization of $\mathop{\rm
tr}\{\nohat{M}\nohat{\rho}(T)\}$ for various inverse temperatures $\beta$
and different damping constants $\gamma_{0}$ in the range typical of
mesoscopic quantum circuits or condensed-phase chemical reactions.
\label{tab:obj}}
\end{table}

\emph{Dynamical cooling.}---Optimal control for closed systems conserves
entropy like any unitary time evolution.  Quantum dissipation
invariably creates mixed states in the subsystem of interest, i.e., if
the initial state is pure the entropy of the open system will
increase. But can optimal control of an \emph{open system} prevent
this or even lower the entropy in other cases? To investigate this
question, we choose the oscillator ground state as target and prepare
both system and environment as thermal states with equal inverse
temperature $\beta =1$. In this symmetric setting, the field $F(t)$
is not needed, since it changes the position, but not the shape of the
wave packet. We therefore consider only the control field $\Delta(t)$ in
the following. The von Neumann entropy of the mixed state is given
by~\cite{holev99}
$ S(\nohat\varrho) = g\left(\sqrt{
\langle q^2\rangle_c \langle p^2\rangle_c - \langle pq+qp \rangle_c^2/4}
\right)$ with
$g(x) = \left(x+\textstyle{\frac{1}{2}}\right)
\log \left(x+\textstyle{\frac{1}{2}}\right)
- \left(x-\textstyle{\frac{1}{2}}\right)
\log \left(x-\textstyle{\frac{1}{2}}\right)$.
\begin{figure}
\includegraphics[width=\linewidth]{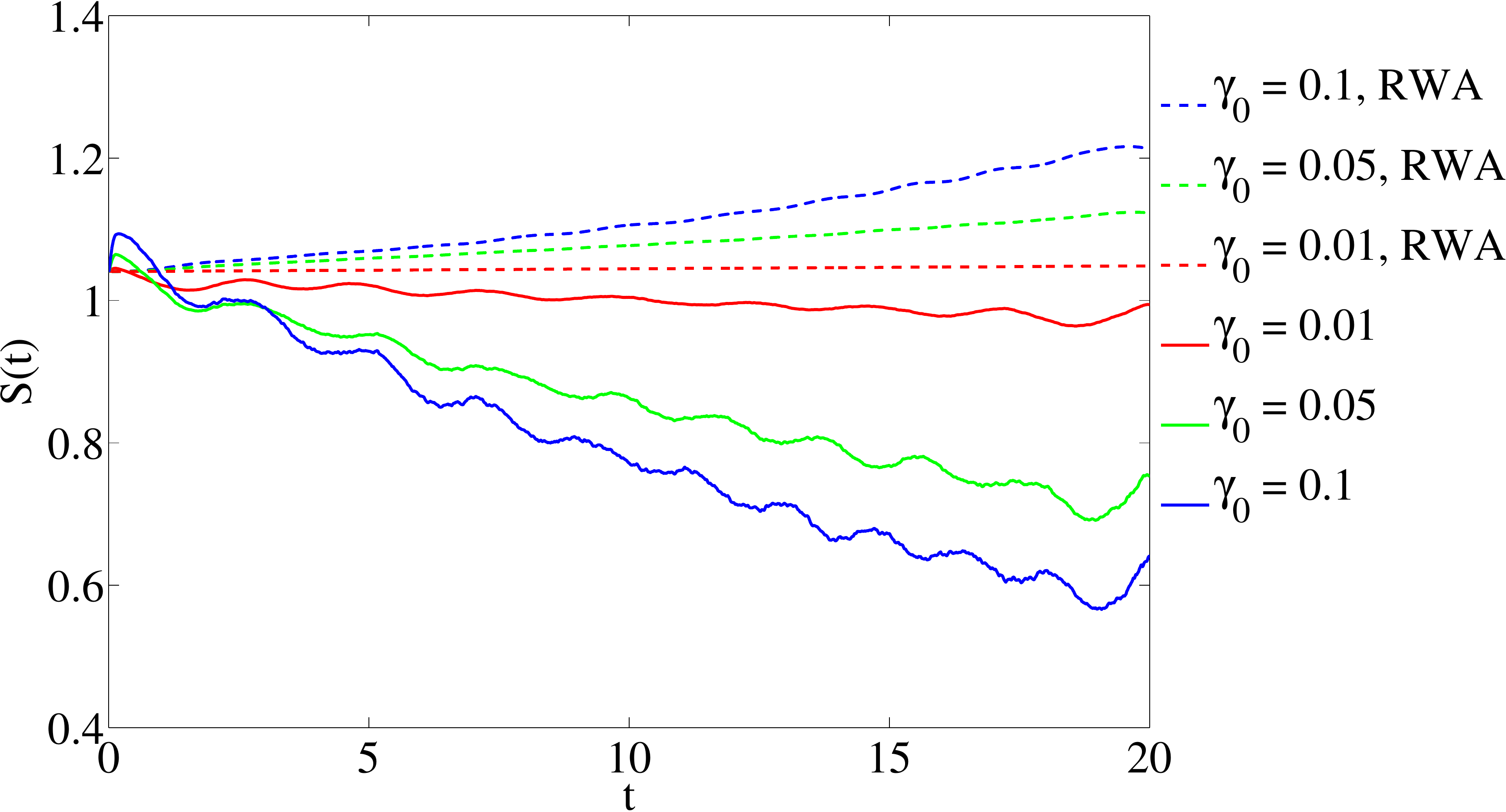}
\caption{(color online) An open quantum system initially equilibrated
with its surroundings loses entropy $S$ under an optimized control
field (solid). In contrast, the standard Markovian/RWA master equation
leads to increased entropy under driving (dashed,
see~\cite{supplprl11}}
\label{fig:entropy}
\end{figure}
We thus obtain the counterintuitive result that a time-dependent
control field can modify dissipative dynamics to the point where its
entropy change turns negative (Fig.\ \ref{fig:entropy}). We attribute
this phenomenon to the \emph{cooperative effect} of driving and
dissipation, since neither of the two alone can cause this.  The
subsystem energy of the final state decreases below its original
thermal value, indicating a \emph{dynamical cooling effect}.  In
contrast, it can be shown (see Supplemental Material
\cite{supplprl11}) that commonly used RWA methods predict heating
\emph{above} the environmental temperature for non-zero
driving. Consistent corrections of master equations for finite
$H_{\mathrm{c}}$ prove to be a formidable challenge~\cite{xu04}.
Moreover, even if $\nohat{H}_{\mathrm{c}}$ could be used in the
construction of the dissipator, the distinction between co- and
counter-rotating terms would hardly be justified. If the control
fields change on the timescale of the reservoir fluctuations, a
`wobbly frame' rather than a rotating frame results.

In contrast to recent proposals for
quantum refrigerators~\cite{pekol07,rey07}, which rely on intricate
band or level structures, we have chosen a model with minimal
structure. The cooling effect found here seems to be a feature of
temporal patterns, not of a specifically designed system.  We also
note that no internal degree of freedom is needed for the effect to
occur.

\emph{Conclusions}.---The present SLN approach to optimal control
enjoys two natural advantages compared to control theory based on
standard Markovian master equations: (i) its noise statistics are by
construction independent of the quantum dynamics, i.e., strong external
driving introduces no need for correction terms, and (ii) one arrives
at the usual state-costate picture required by OCT methods in a
straightforward way. Numerical control of a harmonic degree of freedom
is demonstrated with varying parameters and objectives. Most results
show marked differences compared to the RWA approach, where the
influence of driving on dissipation is neglected.  Efficient
computations are feasible for environmental couplings from weak
damping up to a quality factor as low as $Q\approx 10$. This allows
applications to solid-state devices such as superconducting circuits
with Josephson junctions and condensed-matter phenomena such as
reactive dynamics of small molecules in a solvent or on a
surface. Optimal control of a dissipative quantum system can extract
entropy from a system initially at the same temperature as its
environment. Dynamical cooling in a simple system without special
structural features may be considered as a likely strategy for
mesoscopic quantum refrigeration.

\emph{Acknowledgements}.  We gratefully acknowledge helpful
conversations with S. Montangero and M. Murphy as well as financial
support from Land Baden-W\"{u}rttemberg, DFG (SFB569, SFB/TRR21), EU
(Marie Curie FP7-IEF, AQUTE, DIAMANT, PICC) and Ulm University/UUG.


\bibliographystyle{apsrev}

\bibliography{KrotovJ2,KrotovJ,KrotovDissQSLit}
\cleardoublepage\onecolumngrid


\section*{Supplementary material}

# EPAPS for "Optimal control of open quantum systems: cooperative effects of driving and dissipation"

R. Schmidt,[1] A. Negretti,[2] J. Ankerhold,[1] T. Calarco,[2] and J. T. Stockburger[1]

[1]*Institut für Theoretische Physik, Universität Ulm, Albert-Einstein-Allee 11, 89069 Ulm, Germany.*

[2]*Institut für Quanteninformationsverarbeitung, Universität Ulm, Albert-Einstein-Allee 11, 89069 Ulm, Germany.*

This document serves as a supplement to our article "Optimal control of open quantum systems: cooperative effects of driving and dissipation" [Physical Review Letters (2011)]. It describes the numerical algorithm used, a generalization of the Krotov iterative algorithm to open quantum systems evolving under a stochastic equation of motion. A comparison with the Markovian dynamics of a commonly used quantum master equation shows that the latter cannot reproduce the cooperative cooling effect described in the main text.

## I. OBJECTIVE AND CONSTRAINT

We outline a modification of Krotov's algorithm for the optimization of final-state properties of an open quantum system governed by the stochastic differential equation

$$\dot{\hat{\varrho}}_\xi(t) = \hat{\mathfrak{L}}\hat{\varrho}_\xi(t) := -\frac{i}{\hbar}[\hat{H}_{\rm s}, \hat{\varrho}_\xi(t)] - \frac{i}{\hbar}[\hat{H}_{\rm c}(t), \hat{\varrho}_\xi(t)] - \frac{i}{2\hbar}\gamma_0[\hat{q}, \{\hat{p}, \hat{\varrho}_\xi(t)\}] + \frac{i}{\hbar}\xi(t)[\hat{q}, \hat{\varrho}_\xi(t)], \quad (1)$$

which must be considered as a dynamical constraint in the context of the optimization problem. Here $\gamma_0$ is the damping rate of a Brownian particle in the (ohmic) environment, and $\xi(t)$ is a noise variable whose statistics are governed by the quantum correlation function of the environmental fluctuations.

The objective functionals to be minimized is assumed to depend linearly on the terminal state, i.e., the functional is the expectation value of an observable $\hat{M}$ at the final time $T$, typically a projection operator. The physical terminal state $\hat{\varrho}(T)$ is the stochastic average of the dynamical state $\hat{\varrho}_\xi(T)$, i.e., the objective functional is of the form

$$\mathsf{B}[\mathbf{u}(t)] = {\rm tr}\{\hat{M}\hat{\varrho}(T)\} = \mathbb{E}[{\rm tr}\{\hat{M}\hat{\varrho}_\xi(T)\}] \quad (2)$$

where the vector $\mathbf{u}(t)$ denotes control fields which define the time-dependence of $\hat{H}_{\rm c}(t)$, and the first equality implies the equation of motion (1). We use the symbol $\mathbb{E}$ for the expectation value with respect to the noise $\xi(t)$ only, not for the further average given by tracing over $\hat{\varrho}_\xi(t)$.

## II. EXTENDED OBJECTIVE FUNCTIONAL AND OPTIMAL CONTROL EQUATIONS

Following Refs. [20, 21] (see main text), we use an extended objective functional

$$\mathsf{L}_\xi[\mathbf{u}(t), \hat{\varrho}_\xi(t); \phi_\xi] := \mathsf{G}_\xi(\hat{\varrho}_\xi(T)) - \int_0^T dt\, \mathsf{R}_\xi(t, \mathbf{u}(t), \hat{\varrho}_\xi(t)) - \phi_\xi(0, \hat{\varrho}_\xi(0)), \tag{3}$$

containing the auxiliary, real-valued function $\phi_\xi(t, \hat{\varrho}_\xi(t))$, which depends on the time $t$, the noise realization $\xi(t)$, and the state sample $\hat{\varrho}_\xi(t)$. Further auxiliary functions

$$\mathsf{G}_\xi(\hat{\varrho}_\xi(T)) := \mathrm{tr}\{\hat{M}\hat{\varrho}_\xi(T)\} + \phi_\xi(T, \hat{\varrho}_\xi(T)), \tag{4}$$

$$\mathsf{K}_\xi(t, \mathbf{u}(t), \hat{\varrho}_\xi(t), \,\cdot\,) := \mathrm{tr}\{\,\cdot\,\hat{\mathfrak{L}}\hat{\varrho}_\xi(t)\}, \tag{5}$$

$$\mathsf{R}_\xi(t, \mathbf{u}(t), \hat{\varrho}_\xi(t)) := \mathsf{K}_\xi\left(t, \mathbf{u}(t), \hat{\varrho}_\xi(t), \frac{\partial}{\partial \hat{\varrho}_\xi}\phi_\xi(t, \hat{\varrho}_\xi(t))\right) + \frac{\partial}{\partial t}\phi_\xi(t, \hat{\varrho}_\xi(t)) \tag{6}$$

are chosen such that the expectation value $\mathbb{E}[\mathsf{L}_\xi]$ of the extended functional coincides with the ordinary objective functional for *admissible processes*, i.e., pairs $(\mathbf{u}(t), \{\hat{\varrho}_\xi(t)\})$ of a vector of control fields and a stochastic ensemble $\{\hat{\varrho}_\xi(t)\}$ with the property that the equation of motion (1) is obeyed for the given vector $\mathbf{u}(t)$ and for all noise realizations of $\xi(t)$. In this case the partial derivatives in the definition of $\mathsf{R}_\xi$ combine to form a total derivative, effectively canceling the boundary terms of $\phi_\xi$, contained in Eqs. (3) and (4).

Different choices of $\phi_\xi$ modify the objective functional for non-admissible processes, such as the combination of control fields and states from different iterations. For Krotov's algorithm it is necessary to choose a function $\phi_\xi$ with the following property: For any admissible process, $\mathsf{L}_\xi$ is at a *maximum* with respect to variations of $\hat{\varrho}_\xi$ with the function $\mathbf{u}(t)$ kept fixed, i.e., variations away from the submanifold of admissible processes decrease $\mathsf{L}_\xi$. This means that an appropriate choice of $\phi_\xi$ must conform to the two conditions

$$\mathsf{G}_\xi(\hat{\varrho}_\xi^{(j)}(T)) = \max_{\hat{\varrho}_\xi}\{\mathsf{G}_\xi(\hat{\varrho}_\xi(T))\}, \tag{7}$$

$$\mathsf{R}_\xi(t, \mathbf{u}^{(j)}(t), \hat{\varrho}_\xi^{(j)}(t)) = \min_{\hat{\varrho}_\xi}\{\mathsf{R}_\xi(t, \mathbf{u}^{(j)}(t), \hat{\varrho}_\xi(t))\}, \tag{8}$$

where the superscript $(j)$ denotes the admissible process resulting from an initial guess or a previous iteration.

After making such a choice of $\phi_\xi^{(j)}$, we demand that, in addition to being admissible, the new process $(\mathbf{u}^{(j+1)}(t), \{\hat{\varrho}_\xi^{(j+1)}(t)\})$ is taken from the class of pairs $(\tilde{\mathbf{u}}(t), \{\tilde{\varrho}_\xi(t)\})$ which fulfill the condition

$$\tilde{\mathbf{u}}(t) = \arg\max_{\mathbf{u}(t)} \mathbb{E}\left[\mathsf{R}_\xi(t, \mathbf{u}(t), \tilde{\varrho}_\xi(t))\right] \tag{9}$$

at any time $t$. This condition selects a specific pair $(\mathbf{u}^{(j+1)}(t), \{\hat{\varrho}_\xi^{(j+1)}(t)\})$ from the set of admissible process. Using Eq. (9), the update $\mathbf{u}^{(j)}(t) \to \mathbf{u}^{(j+1)}(t)$ can be performed locally in conjunction with the propagation of the new state $\hat{\varrho}_\xi^{(j+1)}(t)$ since the dependence of the dynamical state on the control fields obeys causality.

What remains to be specified is a mathematical construction which defines specific computational steps using the function $\phi_\xi$. For the linear dynamics of Eq. (1), it turns out that the function $\phi_\xi(t, \hat{\varrho}_\xi)$ appears in the conditions (7) and (8) only in form of its derivative with respect to the state, and since a complete derivative can be formed from them, the *co-state* $\hat{\Lambda}_\xi = \partial \phi_\xi / \partial \hat{\varrho}_\xi$ can be characterized as an additional dynamical variable with the equation of motion

$$\dot{\hat{\Lambda}}_\xi(t) = -\hat{\mathfrak{L}}^\dagger \hat{\Lambda}_\xi(t) = -\frac{i}{\hbar}[\hat{H}_{\rm s}, \hat{\Lambda}_\xi(t)] - \frac{i}{\hbar}[\hat{H}_{\rm c}(t), \hat{\Lambda}_\xi(t)] - \frac{i}{2\hbar}\gamma_0\{\hat{p}, [\hat{q}, \hat{\Lambda}_\xi(t)]\} + \frac{i}{\hbar}\xi(t)[\hat{q}, \hat{\Lambda}_\xi(t)]. \quad (10)$$

Solving this equation with the end-time boundary condition $\hat{\Lambda}_\xi(T) + \hat{M} = 0$, resulting from (7), amounts to a (fully sufficient) construction of $\phi_\xi$ to first order in variations of $\hat{\varrho}_\xi$ around $\hat{\varrho}_\xi^{(j)}$.

In a variant of the algorithm, described by Sklarz and Tannor [21], the implicit condition (9) is substituted by an explicit form

$$u_k^{(j+1)}(t) = u_k^{(j)}(t) + \frac{1}{\lambda_k(t)} \mathbb{E}\left[{\rm tr}\left\{\hat{\Lambda}_\xi^{(j)}(t) \frac{\partial \hat{\mathfrak{L}}}{\partial u_k} \hat{\varrho}_\xi^{(j+1)}(t)\right\}\right], \quad (11)$$

which is equivalent to Eq. (9) if a suitable ($j$-dependent) cost functional is added to B. It is to be noted that the simultaneous appearance of $j$ and $j+1$ on the r.h.s. of this equation gives this update law properties quite different from a simple gradient search. The functions $\lambda_k(t)$ arising from this cost functional may be adjusted to tune convergence properties. In summary, the algorithm works as follows:

1. An initial guess for the control fields $\mathbf{u}^{(0)}(t)$ is chosen.

2. The equation of motion (1) with $\mathbf{u}^{(0)}(t)$ is solved with initial condition $\hat{\varrho}(0)$ for an ensemble of trajectories $\{\hat{\varrho}_\xi(t)\}$.

3. The co-state ensemble is initialized using the condition $\hat{\Lambda}_\xi(T) + \hat{M} = 0$ and then propagated backwards to the initial time with Eq. (10).

4. The ensembles $\{\hat{\varrho}_\xi(t)\}$ and $\{\hat{\Lambda}_\xi(t)\}$ are propagated forward until the end time $T$, but at each time step the co-states are propagated with the old control fields, that is, the ones obtained from the $j$-th iteration, whereas the states $\hat{\varrho}_\xi(t)$ are propagated with the new control fields given by the update law (11). (Alternatively, stored values from step 3 can be used).

5. The objective functional (2) is computed, and steps 3 and 4 are repeated until the value of objective functional is in the desired range.

This optimization algorithm applies not only to Eq. (1), but to any type of linear dynamics subject to additive or multiplicative noise.

### III. PROOF OF MONOTONIC CONVERGENCE

The procedure outlined above guarantees that each step of the iteration reduces the objective functional, that is, $\mathbb{E}[\mathsf{B}_\xi[\mathbf{u}^{(j+1)}(t), \hat{\varrho}_\xi^{(j+1)}(t)]] \leq \mathbb{E}[\mathsf{B}_\xi[\mathbf{u}^{(j)}(t), \hat{\varrho}_\xi^{(j)}(t)]]$. This becomes evident if the change of the objective functional is decomposed into three terms,

$$\mathbb{E}[\mathsf{L}_\xi[\mathbf{u}^{(j)}(t), \hat{\varrho}_\xi^{(j)}(t), \phi_\xi]] - \mathbb{E}[\mathsf{L}_\xi[\mathbf{u}^{(j+1)}(t), \hat{\varrho}_\xi^{(j+1)}(t), \phi_\xi]] = \Delta_1 + \Delta_2 + \Delta_3, \tag{12}$$

where

$$\Delta_1 = \int_0^T dt\, \mathbb{E}\left[\mathsf{R}(t, \mathbf{u}^{(j)}(t), \hat{\varrho}_\xi^{(j+1)}(t)) - \mathsf{R}(t, \mathbf{u}^{(j)}(t), \hat{\varrho}_\xi^{(j)}(t))\right] \tag{13}$$

$$\Delta_2 = \int_0^T dt\, \mathbb{E}\left[\mathsf{R}(t, \mathbf{u}^{(j+1)}(t), \hat{\varrho}_\xi^{(j+1)}(t)) - \mathsf{R}(t, \mathbf{u}^{(j)}(t), \hat{\varrho}_\xi^{(j+1)}(t))\right] \tag{14}$$

$$\Delta_3 = \mathbb{E}\left[\mathsf{G}(\hat{\varrho}_\xi^{(j)}(T)) - \mathsf{G}(\hat{\varrho}_\xi^{(j+1)}(T))\right] \tag{15}$$

Now $\Delta_1 \geq 0$ and $\Delta_3 \geq 0$ follow from Eq. (8) and Eq. (7), respectively, whereas $\Delta_2 \geq 0$ is a property of the update law (9). We have thus demonstrated monotonicity for the modified algorithm, which ensures convergence for an arbitrary initial guess with any objective functional bounded from below.

Recent work by S. Schirmer and P. de Fouquieres [New J. Phys. **13**, 073029 (2011)] points out that proofs of monotonicity for Krotov-type iteration methods are not rigorous if time-discrete approximations are substituted for the dynamics of state and co-state. For all practical purposes, however, the typical benefits of the Krotov method persist, i.e., convergence for a wide range of initial guesses and rapid improvement during the first iteration steps.

### IV. DRIVEN DYNAMICS WITH FIXED RWA DISSIPATOR: EFFECTIVE RESERVOIR

For comparison, we study the control problem of the parametrically driven harmonic oscillator, modeling the open system dynamics through a Lindblad Master equation. From a system-reservoir model, this equation results from performing the Born, Markov, and RWA approximations on the

system-reservoir interactions. In the dimensionless units introduced in the main text, the Master equation reads

$$\begin{aligned}
\frac{d}{dt}\hat{\rho} &= -i[\hat{H}(t), \hat{\rho}] \\
&+ \gamma_0(N_\beta + 1)\left(\hat{a}\hat{\rho}\hat{a}^\dagger - \tfrac{1}{2}\{\hat{a}^\dagger\hat{a}, \hat{\rho}\}\right) \\
&+ \gamma_0 N_\beta \left(\hat{a}^\dagger\hat{\rho}\hat{a} - \tfrac{1}{2}\{\hat{a}\hat{a}^\dagger, \hat{\rho}\}\right)
\end{aligned} \quad (16)$$

with

$$\hat{H}(t) = \hat{a}^\dagger\hat{a} + \frac{1}{2} + \frac{\Delta(t)}{4}(\hat{a} + \hat{a}^\dagger)^2 = \left(1 + \frac{\Delta(t)}{2}\right)(\hat{a}^\dagger\hat{a} + \tfrac{1}{2}) + \frac{\Delta(t)}{4}(\hat{a}^2 + \hat{a}^{\dagger 2}) \,. \quad (17)$$

Following common practice (see, however, Refs. [10, 11]), we have defined the rotating frame of the RWA through the time-independent Hamiltonian $\hat{H}_0 = \hat{a}^\dagger\hat{a} + \tfrac{1}{2}$, not the full Hamiltonian $\hat{H}(t)$. Accordingly, the dissipative terms of Eq. (16) contain the thermal occupation number $N_\beta = (e^\beta - 1)^{-1}$ of the unperturbed oscillator.

Due to the mismatch between the Hamiltonians of the pre-set rotating frame and of the driven dynamics, oscillatory terms reappear when the dissipative terms are transformed to the interaction picture. In order to show this, we define the interaction picture with respect to $\hat{H}(t)$,

$$\hat{\rho}(t) = \hat{U}(t)\hat{\rho}_{\mathrm{I}}(t)\hat{U}^\dagger(t) \,, \quad \dot{\hat{U}} = i\hat{H}\hat{U} \,. \quad (18)$$

The Master equation for $\hat{\rho}_{\mathrm{I}}$ then reads

$$\begin{aligned}
\frac{d}{dt}\hat{\rho}_{\mathrm{I}} &= \gamma_0(N_\beta + 1)\left(\hat{b}\hat{\rho}_{\mathrm{I}}\hat{b}^\dagger - \tfrac{1}{2}\{\hat{b}^\dagger\hat{b}, \hat{\rho}_{\mathrm{I}}\}\right) \\
&+ \gamma_0 N_\beta \left(\hat{b}^\dagger\hat{\rho}_{\mathrm{I}}\hat{b} - \tfrac{1}{2}\{\hat{b}\hat{b}^\dagger, \hat{\rho}_{\mathrm{I}}\}\right) \,,
\end{aligned} \quad (19)$$

where $\hat{b} = \hat{U}^\dagger \hat{a} \hat{U}$ and its adjoint are time dependent, following the equations of motion

$$\dot{\hat{b}} = -i\left(1 + \tfrac{\Delta}{2}\right)\hat{b} - i\tfrac{\Delta}{2}\hat{b}^\dagger \quad (20)$$

$$\dot{\hat{b}}^\dagger = i\tfrac{\Delta}{2}\hat{b} + i\left(1 + \tfrac{\Delta}{2}\right)\hat{b}^\dagger \,. \quad (21)$$

From this, one easily concludes that $b$ is a linear combination of $\hat{a}$ and $\hat{a}^\dagger$ with $[\hat{b}, \hat{b}^\dagger] = 1$, i.e., the map $\hat{a} \mapsto \hat{b}$ is a Bogoliubov transformation

$$\hat{b} = e^{i\theta} \cosh y \, \hat{a} + e^{i\varphi} \sinh y \, \hat{a}^\dagger \quad (22)$$

with real (and, in our case, time-dependent) parameters $\theta$, $\phi$, and $y$. Using this equation, we obtain

the form

$$\begin{aligned}
\frac{d}{dt}\hat{\rho}_\mathrm{I} &= \gamma_0(N_\beta+1)\cosh^2 y \left(\hat{a}\hat{\rho}_\mathrm{I}\hat{a}^\dagger - \tfrac{1}{2}\{\hat{a}^\dagger\hat{a},\hat{\rho}_\mathrm{I}\}\right) \\
&+ \gamma_0(N_\beta+1)\sinh^2 y \left(\hat{a}^\dagger\hat{\rho}_\mathrm{I}\hat{a} - \tfrac{1}{2}\{\hat{a}\hat{a}^\dagger,\hat{\rho}_\mathrm{I}\}\right) \\
&+ \gamma_0 N_\beta \cosh^2 y \left(\hat{a}^\dagger\hat{\rho}_\mathrm{I}\hat{a} - \tfrac{1}{2}\{\hat{a}\hat{a}^\dagger,\hat{\rho}_\mathrm{I}\}\right) \\
&+ \gamma_0 N_\beta \sinh^2 y \left(\hat{a}\hat{\rho}_\mathrm{I}\hat{a}^\dagger - \tfrac{1}{2}\{\hat{a}^\dagger\hat{a},\hat{\rho}_\mathrm{I}\}\right) \\
&+ \gamma_0(2N_\beta+1)\sinh y \cosh y\, e^{i(\theta-\varphi)} \left(\hat{a}\hat{\rho}_\mathrm{I}\hat{a} - \tfrac{1}{2}\{\hat{a}^2,\hat{\rho}_\mathrm{I}\}\right) \\
&+ \gamma_0(2N_\beta+1)\sinh y \cosh y\, e^{-i(\theta-\varphi)} \left(\hat{a}^\dagger\hat{\rho}_\mathrm{I}\hat{a}^\dagger - \tfrac{1}{2}\{\hat{a}^{\dagger 2},\hat{\rho}_\mathrm{I}\}\right)
\end{aligned} \quad (23)$$

for the Master equation, where all of the prefactors depend on time through $\theta$, $\varphi$, and $y$. Typical solutions lead to oscillating prefactors, meaning that Eq. (23) can be further simplified through time coarse graining. This in itself can be seen as evidence that the initial application of the RWA was not consistent. However, coarse-graining at this point is appropriate if one wants to compare the features of the dynamics predicted by Eqs. (16) and (23) with the exact dynamics. The first four terms of Eq. (23) contain functions of time which are squares of real time-dependent functions, therefore they typically dominate the result of the coarse graining procedure. Collecting the dominant terms results in the simplified Master equation

$$\begin{aligned}
\frac{d}{dt}\hat{\rho}_\mathrm{I} &= \gamma_0(\tilde{N}+1)\left(\hat{a}\hat{\rho}_\mathrm{I}\hat{a}^\dagger - \tfrac{1}{2}\{\hat{a}^\dagger\hat{a},\hat{\rho}_\mathrm{I}\}\right) \\
&+ \gamma_0\tilde{N}\left(\hat{a}^\dagger\hat{\rho}_\mathrm{I}\hat{a} - \tfrac{1}{2}\{\hat{a}\hat{a}^\dagger,\hat{\rho}_\mathrm{I}\}\right)
\end{aligned} \quad (24)$$

with

$$2\tilde{N}+1 = (2N_\beta+1)\overline{\cosh(2y)} \geq 2N_\beta+1 , \quad (25)$$

where the bar indicates time coarse graining. The immediate interpretation of this result is obvious: Up to a unitary transform, the dynamics of Eqs. (16) and (23) effectively describes relaxation towards a thermal oscillator state with a higher-than-thermal occupation number $\tilde{N}$. This state therefore has a higher entropy than the thermal state obtained by equilibrating without driving. Accordingly, with the dynamics approximated by Eq. (16), optimal control cannot find solutions which push the system entropy below its equilibrium value. This is consistent with our numerical findings, in particular the observation that optimal control based on Eq. (16), with overlap to the ground state as objective, converges to the trivial result $\Delta \equiv 0$. The alluringly simple method of constructing the dissipator based on a Hamiltonian without driving is an oversimplification which fails to reproduce cooperative effects of driving and dissipation, most notably, the cooling effect reported in the main text.

%
\end{document}